%
%

\input harvmac

\def\C{{\bf C}}

\def\Z{{\bf Z}}

\def\a{\alpha}

\def\e{\epsilon}

\def\l{\lambda}
\def\L{\Lambda}
\def\m{\mu}
\def\n{\nu}
\def\r{\rho}

\def\F{\Phi}
\def\w{\omega}

\def\v{\varphi}

\def\P{{\bf P}}

\def\d{\partial}

\def\Tr{{\rm Tr}}
\def\hf{{1\over 2}}

\def\cC{{\cal C}}
\def\cF{{\cal F}}
\def\cN{{\cal N}}
\def\cW{{\cal W}}

\def\({\bigl(}
\def\){\bigr)}
\def\<{\langle\,}
\def\>{\,\rangle}

\lref\thooft{
G.~'t Hooft, ``A Planar Diagram Theory For Strong Interactions,''
Nucl.\ Phys.\ B {\bf 72}, 461 (1974).}

\lref\mm{
P.~Ginsparg and G.~W.~Moore,
``Lectures On 2-D Gravity And 2-D String Theory,''
arXiv:hep-th/9304011.}

\lref\mmm{
P.~Di Francesco, P.~Ginsparg and J.~Zinn-Justin,
``2-D Gravity and random matrices,''
Phys.\ Rept.\  {\bf 254}, 1 (1995)
[arXiv:hep-th/9306153]}

\lref\kontsevich{M.~Kontsevich,
``Intersection Theory On The Moduli Space Of Curves And The Matrix
Airy Function,'' Commun.\ Math.\ Phys.\ {\bf 147}, 1 (1992).}

\lref\wittentop{E.~Witten,
``On The Structure Of The Topological Phase Of Two-Dimensional
Gravity,'' Nucl.\ Phys.\ B {\bf 340}, 281 (1990)}

\lref\matrix{T.~Banks, W.~Fischler, S.~H.~Shenker and L.~Susskind,
``M theory as a matrix model: A conjecture,''
Phys.\ Rev.\ D {\bf 55}, 5112 (1997)
[arXiv:hep-th/9610043].}

\lref\adscft{
O.~Aharony, S.~S.~Gubser, J.~M.~Maldacena, H.~Ooguri and Y.~Oz,
``Large N field theories, string theory and gravity,''
Phys.\ Rept.\  {\bf 323}, 183 (2000)
[arXiv:hep-th/9905111].
}

\lref\ghv{D. Ghoshal and C. Vafa, ``c=1 string as the topological theory
of the conifold,''
Nucl.\ Phys.\ B {\bf 453}, 121 (1995)
[arXiv:hep-th/9506122].}

\lref\klmkv{S. Kachru, A. Klemm, W. Lerche, P. Mayr, C. Vafa,
``Nonperturbative Results on the Point Particle Limit of N=2 Heterotic String
    Compactifications,'' Nucl.\ Phys.\ B {\bf 459}, 537 (1996)
[arXiv:hep-th/9508155].}

\lref\klw{A. Klemm, W. Lerche, P. Mayr, C.Vafa, N. Warner,
``Self-Dual Strings and N=2 Supersymmetric Field Theory,''
 Nucl.\ Phys. \ B {\bf 477}, 746 (1996)
[arXiv:hep-th/9604034].}

\lref\kmv{S. Katz, P. Mayr, C. Vafa,
``Mirror symmetry and Exact Solution of 4D N=2 Gauge Theories I,''
Adv.\ Theor.\ Math.\ Phys. {\bf 1}, 53 (1998)
[arXiv:hep-th/9706110].}

\lref\gv{R.~Gopakumar and C.~Vafa,
``On the gauge theory/geometry correspondence,''
Adv.\ Theor.\ Math.\ Phys.\  {\bf 3}, 1415 (1999)
[arXiv:hep-th/9811131].
}

\lref\edel{J.D. Edelstein, K. Oh and R. Tatar, ``Orientifold,
geometric transition and large $N$ duality for SO/Sp gauge theories,''
JHEP {\bf 0105}, 009 (2001)
[arXiv:hep-th/0104037].}

\lref\dasg{K. Dasgupta, K. Oh and R. Tatar, ``Geometric transition, large
$N$ dualities and MQCD dynamics,''  Nucl. Phys.
B {\bf 610}, 331 (2001)
[arXiv:hep-th/0105066]\semi -----,
``Open/closed string dualities and Seiberg
duality from geometric transitions in M-theory,''
[arXiv:hep-th/0106040]\semi -----, ``Geometric transition versus cascading
solution,''
JHEP {\bf 0201},  031 (2002)
[arXiv:hep-th/0110050].}

\lref\hv{K. Hori and C. Vafa,
``Mirror Symmetry,''
[arXiv:hep-th/0002222].}

\lref\hiv{K. Hori, A. Iqbal and C. Vafa,
``D-Branes And Mirror Symmetry,''
[arXiv:hep-th/0005247].}

\lref\vaug{C.~Vafa,
``Superstrings and topological strings at large N,''
J.\ Math.\ Phys.\  {\bf 42}, 2798 (2001)
[arXiv:hep-th/0008142].}

\lref\civ{
F.~Cachazo, K.~A.~Intriligator and C.~Vafa,
``A large $N$ duality via a geometric transition,''
Nucl.\ Phys.\ B {\bf 603}, 3 (2001)
[arXiv:hep-th/0103067].}

\lref\ckv{
F.~Cachazo, S.~Katz and C.~Vafa,
``Geometric transitions and $N = 1$ quiver theories,''
arXiv:hep-th/0108120.}

\lref\cfikv{
F.~Cachazo, B.~Fiol, K.~A.~Intriligator, S.~Katz and C.~Vafa,
``A geometric unification of dualities,'' Nucl.\ Phys.\ B {\bf 628}, 3
(2002) [arXiv:hep-th/0110028].}

\lref\cv{
F.~Cachazo and C.~Vafa, ``$N=1$ and $N=2$ geometry from fluxes,''
arXiv:hep-th/0206017.}

\lref\ov{
H.~Ooguri and C.~Vafa, ``Worldsheet derivation of a large $N$ duality,''
arXiv:hep-th/0205297.}

\lref\av{
M.~Aganagic and C.~Vafa, ``$G_2$ manifolds, mirror symmetry,
and geometric engineering,'' arXiv:hep-th/0110171.}

\lref\digra{
D.~E.~Diaconescu, B.~Florea and A.~Grassi, ``Geometric transitions and
open string instantons,'' arXiv:hep-th/0205234.}

\lref\amv{
M.~Aganagic, M.~Marino and C.~Vafa,
``All loop topological string amplitudes from Chern-Simons theory,''
arXiv:hep-th/0206164.}

\lref\dfg{
D.~E.~Diaconescu, B.~Florea and A.~Grassi, ``Geometric transitions,
del Pezzo surfaces and open string instantons,''
arXiv:hep-th/0206163.}

\lref\kkl{
S.~Kachru, S.~Katz, A.~E.~Lawrence and J.~McGreevy,
``Open string instantons and superpotentials,''
Phys.\ Rev.\ D {\bf 62}, 026001 (2000)
[arXiv:hep-th/9912151].}

\lref\bcov{
M.~Bershadsky, S.~Cecotti, H.~Ooguri and C.~Vafa, ``Kodaira-Spencer
theory of gravity and exact results for quantum string amplitudes,''
Commun.\ Math.\ Phys.\ {\bf 165}, 311 (1994) [arXiv:hep-th/9309140].
}

\lref\witcs{
E.~Witten,
``Chern-Simons gauge theory as a string theory,''
arXiv:hep-th/9207094.
}

\lref\naret{I. Antoniadis, E. Gava, K.S. Narain, T.R. Taylor,
``Topological Amplitudes in String Theory,''
Nucl.\ Phys.\ B\ {\bf 413}, 162 (1994)
[arXiv:hep-th/9307158].}

\lref\witf{E. Witten,
``Solutions Of Four-Dimensional Field Theories Via M Theory,''
Nucl.\ Phys.\ B\ {\bf 500}, 3 (1997)
[arXiv:hep-th/9703166].}

\lref\shenker{
S.~H.~Shenker, ``The Strength Of Nonperturbative Effects In String
Theory,'' in Proceedings Cargese 1990, {\it Random surfaces and
quantum gravity}, 191--200.  }

\lref\berwa{
M.~Bershadsky, W.~Lerche, D.~Nemeschansky and
N.~P.~Warner, ``Extended $N=2$ superconformal structure of gravity and W
gravity coupled to matter,'' Nucl.\ Phys.\ B {\bf 401}, 304 (1993)
[arXiv:hep-th/9211040]}

\lref\loop{
G.~Akemann, ``Higher genus correlators for the Hermitian matrix model
with multiple cuts,'' Nucl.\ Phys.\ B {\bf 482}, 403 (1996)
[arXiv:hep-th/9606004].  }

\lref\wiegmann{P.B. Wiegmann and A. Zabrodin, ``Conformal maps
and integrable hierarchies,'' arXiv:hep-th/9909147.}

\lref\cone{R.~Dijkgraaf and C.~Vafa, to appear.}

\lref\kazakov{
S.~Y.~Alexandrov, V.~A.~Kazakov and I.~K.~Kostov,
``Time-dependent backgrounds of 2D string theory,''
arXiv:hep-th/0205079.}

\lref\dj{
S.~R.~Das and A.~Jevicki,
``String Field Theory And Physical Interpretation Of $D=1$ Strings,''
Mod.\ Phys.\ Lett.\ A {\bf 5}, 1639 (1990).}

\lref\givental{
A.B.~Givental, ``Gromov-Witten invariants and quantization of quadratic hamiltonians,''
 arXiv:math.AG/0108100.}

\lref\op{
A.~Okounkov and R.~Pandharipande, ``Gromov-Witten theory, Hurwitz
theory, and completed cycle,'' arXiv:math.AG/0204305.}

\lref\dijk{ R.~Dijkgraaf, ``Intersection theory, integrable hierarchies and
topological field theory,'' in Cargese Summer School on {\it New Symmetry
Principles in Quantum Field Theory} 1991,
[arXiv:hep-th/9201003].}

\lref\gw{
D.~J.~Gross and E.~Witten, ``Possible Third Order Phase Transition In
The Large $N$ Lattice Gauge Theory,'' Phys.\ Rev.\ D {\bf 21}, 446
(1980).}

\lref\sw{
N.~Seiberg and E.~Witten, ``Electric-magnetic duality, monopole
condensation, and confinement in $N=2$ supersymmetric Yang-Mills
theory,'' Nucl.\ Phys.\ B {\bf 426}, 19 (1994) [Erratum-ibid.\ B {\bf
430}, 485 (1994)] [arXiv:hep-th/9407087].}

\lref\kkv{
S.~Katz, A.~Klemm and C.~Vafa, ``Geometric engineering of quantum
field theories,'' Nucl.\ Phys.\ B {\bf 497}, 173 (1997)
[arXiv:hep-th/9609239].}

\lref\taylor{
W.~I.~Taylor, ``D-brane field theory on compact spaces,'' Phys.\
Lett.\ B {\bf 394}, 283 (1997) [arXiv:hep-th/9611042].}

\lref\kmmms{
S.~Kharchev, A.~Marshakov, A.~Mironov, A.~Morozov and S.~Pakuliak,
``Conformal matrix models as an alternative to conventional
multimatrix models,'' Nucl.\ Phys.\ B {\bf 404}, 717 (1993)
[arXiv:hep-th/9208044].}

\lref\dv{
R.~Dijkgraaf and C.~Vafa, ``Matrix models, topological strings, and
supersymmetric gauge theories,'' arXiv:hep-th/0206255.}

\lref\kostov{
I.~K.~Kostov,
``Gauge invariant matrix model for the A-D-E closed strings,''
Phys.\ Lett.\ B {\bf 297}, 74 (1992)
[arXiv:hep-th/9208053].
}

\lref\ot{
K.~h.~Oh and R.~Tatar, ``Duality and confinement in $N=1$
supersymmetric theories from geometric transitions,''
arXiv:hep-th/0112040.}

\lref\ovknot{
H.~Ooguri and C.~Vafa,
``Knot invariants and topological strings,''
Nucl.\ Phys.\ B {\bf 577}, 419 (2000)
[arXiv:hep-th/9912123].
}

\lref\dvv{
R.~Dijkgraaf, H.~Verlinde and E.~Verlinde, ``Loop Equations
And Virasoro Constraints In Nonperturbative 2-D Quantum Gravity,''
Nucl.\ Phys.\ B {\bf 348}, 435 (1991).}

\lref\kawai{
M.~Fukuma, H.~Kawai and R.~Nakayama, ``Continuum Schwinger-Dyson
Equations And Universal Structures In Two-Dimensional Quantum
Gravity,'' Int.\ J.\ Mod.\ Phys.\ A {\bf 6}, 1385 (1991).}

\lref\nekrasov{
N.~A.~Nekrasov, ``Seiberg-Witten prepotential from instanton
counting,'' arXiv:hep-th/0206161.}

\lref\ki{
I.~K.~Kostov, ``Bilinear functional equations in 2D quantum gravity,''
in Razlog 1995, {\it New trends in quantum field theory}, 77--90,
[arXiv:hep-th/9602117].}

\lref\kii{
I.~K.~Kostov, ``Conformal field theory techniques in random matrix
models,'' arXiv:hep-th/9907060.}

\lref\morozov{
A.~Morozov, ``Integrability And Matrix Models,'' Phys.\ Usp.\ {\bf
37}, 1 (1994) [arXiv:hep-th/9303139].}

\lref\fo{
H.~Fuji and Y.~Ookouchi, ``Confining phase superpotentials for SO/Sp
gauge theories via geometric transition,'' arXiv:hep-th/0205301.
}

\lref\marinor{M. Marino, ``Chern-Simons theory, matrix integrals, and
perturbative three-manifold invariants,''
[arXiv:hep-th/0207096].}

\Title
 {\vbox{
 \hbox{hep-th/0207106}
 \hbox{HUTP-02/A030}
 \hbox{ITFA-2002-24}
}}
{\vbox{
\centerline{On Geometry and Matrix Models}
}}
\bigskip
\centerline{Robbert Dijkgraaf}
\vskip.05in
\centerline{\sl Institute for Theoretical Physics \&}
\centerline{\sl Korteweg-de Vries Institute for Mathematics}
\centerline{\sl University of Amsterdam}
\centerline{\sl 1018 TV Amsterdam, The Netherlands }
\smallskip
\centerline{and}
\smallskip
\centerline{Cumrun Vafa}
\vskip.05in
\centerline{\sl Jefferson Physical Laboratory}
\centerline{\sl Harvard University}
\centerline{\sl Cambridge, MA 02138, USA}

\vskip .1in\centerline{\bf Abstract}

\smallskip

We point out two extensions of the relation between matrix models,
topological strings and $\cN=1$ supersymmetric gauge theories. First,
we note that by considering double scaling limits of {\it unitary}
matrix models one can obtain large $N$ duals of the local Calabi-Yau
geometries that engineer $\cN=2$ gauge theories. In particular, a
double scaling limit of the Gross-Witten one-plaquette lattice model
gives the $SU(2)$ Seiberg-Witten solution, including its induced
gravitational corrections.  Secondly, we point out that the effective
superpotential terms for $\cN=1$ $ADE$ quiver gauge theories is
similarly computed by large $N$ multi-matrix models, that have been
considered in the context of $ADE$ minimal models on random
surfaces. The associated spectral curves are multiple branched covers
obtained as Virasoro and $W$-constraints of the partition function.

\Date{July, 2002}


\newsec{Introduction}

In \dv\ we have shown how the effective superpotential in $\cN=1$
supersymmetric gauge theories, that are obtained by breaking an
$\cN=2$ super Yang-Mills theory by adding a tree-level superpotential
$W(\F)$ for the adjoint scalar $\F$, can be computed by a large $N$
hermitian matrix models. More precisely, the effective superpotential
of the $\cN=1$ theory considered as a function of the gluino
condensates $S_i$ is---apart from the universal $S_i\log (S_i/\L)$
terms coming from the pure $\cN=1$ Yang-Mills theory---given {\it
exactly} by a perturbative series that is computed by the planar
diagrams of the matrix model with potential $W(\F)$. Furthermore, the
contributions of this gauge theory to the induced supergravity
corrections $R^2F^{2g-2}$ (with $R$ the Riemann curvature and $F$ the
graviphoton field strength) are similarly computed exactly by the
genus $g>0$ matrix diagrams.

This gauge theory/matrix model correspondence was a consequence of the
large $N$ dualities of \refs{\gv,\vaug ,\civ } that relate the
computation of holomorphic F-terms in the world-volume theories of
D-branes to partition functions of topological strings in local
Calabi-Yau geometries---a relation that was further explored in
\refs{\edel,\dasg,\ckv,\cfikv,\ot,\fo,\cv}.
In the simplest case these local non-compact Calabi-Yau manifolds take
the form
$$
vv'+ y^2-W'(x)^2+f(x)=0.
$$
One finds that in the B-model topological string the tree-level free
energy can be computed in terms of the periods of the meromorphic
differential $ydx$ on the associated Riemann surface
$$
y^2-W'(x)^2+f(x)=0.
$$

As we argued in \dv\ this curve and the associated special geometry
arises naturally from the large $N$ dynamics of the matrix integral
with action $W(\F)$. But we should stress again that the relation
with matrix models goes beyond the planar limit. The higher genus
string partition functions $\cF_g$ and the related gravitational
couplings of the gauge theories are exactly computed in the $1/N$
expansion of the matrix models.

Let us briefly summarize these connections, for more details see
\dv. We start with the hermitian matrix integral
$$
\int d\F\cdot e^{-S(\F)}
$$
with action
$$
S(\F)= {1\over g_s}\Tr\,W(\F),
$$
and $W(x)$ is a polynomial of degree $n+1$.  The matrix integral can
be reduced to an integral over the eigenvalues $x_1,\ldots,x_N$ of
$\F$ in the potential $W(x)$. In the classical limit $g_s\to 0$, where
one ignores the interactions among the eigenvalues, the equation of
motion is given by
\eqn\hj{
y(x)=g_s {\d S\over \d x}=W'(x)=0.
}
The associated classical spectral curve is
\eqn\clas{
y^2-W'(x)^2=0,
}
where $x,y$ can be considered as complex variables.  Writing
$W'(x)=\prod_i (x-a_i)$ we see that this singular genus zero planar
curve has $n$ double points at the critical points $x=a_i$.

Sometimes it can be helpful to think of the $(x,y)$-plane as a phase
space, with $y$ the momentum conjugate to $x$, as given by the
Hamilton-Jacobi equation \hj. Then \clas\ has an interpretation as the
zero-energy level set of the (bosonic part of the) supersymmetric
quantum mechanics Hamiltonian associated to the superpotential $W(x)$,
and $S(x)$ can be thought of as the semi-classical WKB action of the
associated quantum mechanical ground state $\Psi(x)\sim e^{-S(x)}$.

Classically, the $N$ eigenvalues will cluster in groups of $N_i$ in
the critical points $a_i$ where they will form some meta-stable
state. The relative number of eigenvalues or filling fraction of the
critical point $a_i$ we will denote as
$$
\nu_i=N_i/N.
$$
If $g_s$ is not zero, we have to take into account the Coulomb
interaction that results from integrating out the angular,
off-diagonal components of the matrix $\F$. The equation of motion of
a single eigenvalue $x$ in the presence of the Dyson gas of
eigenvalues $x_1,\ldots,x_N$ is now modified to
$$
y=W'(x)-2g_s \sum_{I=1}^N {1\over x-x_I}.
$$

We will now take the large $N$ 't Hooft limit keeping both $\mu=g_sN$
and the filling fractions $\nu_i$ fixed. In this case each critical
point has its own 't Hooft coupling
$$
\mu_i=g_s N_i = \mu\,\nu_i.
$$
The collective dynamics of these eigenvalues in the large $N$ limit can be
summarized geometrically as follows. Each of the $n$ double points
$x=a_i$ gets resolved into two branch points $a_i^+,a_i^-$. The
resulting branch cuts $A_i=[a_i^-,a_i^+]$ are filled by a continuous
density of eigenvalues that behave as fermions and spread out due to
the Pauli exclusion principle.  This process of splitting up of double
points is very analogous to transition from the classical to the
quantum moduli space in the Seiberg-Witten solution of $\cN=2$
supersymmetric gauge theories \sw---a relation that was explained
in \cv . The resolution of double points is captured by
deforming the classical spectral curve \clas\ into the quantum curve
\eqn\quantum{
y^2-W'(x)^2+ \mu f(x)=0,
}
where the quantum deformation $f(x)$ is a polynomial of degree
$n-1$. The filling fraction $\nu_i$ is related to the size of the
branch cut $A_i$. Roughly, the higher the proportion of eigenvalue at
the critical point $a_i$, the larger the cut. More precisely, we have
$$
\mu_i =\mu\,\nu_i = {1\over 2\pi i} \oint_{A_i} y(x) dx
$$
(This equation can be considered as analogous to the Bohr-Sommerfeld
quantization condition.) Finally the tree-level free energy
$\cF(\mu_i)$ can be computed in terms of the dual $B$-periods by the
special geometry relations
$$
{\d\cF \over \d\mu_i} = \oint_{B_i} y(x)dx,
$$
where the cycles $B_i$ run from the branch cuts to some cut-off point
at infinity.

\newsec{Unitary matrix models and the Seiberg-Witten solution}

We have reviewed how the effective superpotential in $\cN=1$
supersymmetric gauge theories obtained by breaking an $\cN=2$ super
Yang-Mills theory can be computed by a large $N$ hermitian matrix
model. This raises the question whether there exists a matrix model
that computes {\it directly} holomorphic F-terms in the underlying
undeformed $\cN=2$ theory as described by the rigid special geometry
of the Seiberg-Witten solution \sw. As explained in \cv\ it is indeed
possible to extract the geometry of the $\cN=2$ solution from the
effective superpotential of the $\cN=1$ deformation of $\cN=2$ theory
and thereby, indirectly, from the associated matrix model. But here we
will pursue a more direct relation---we will show that indeed there is
a large $N$ matrix model that computes the SW solution directly. The
matrix model for $\cN=2$ supersymmetric $SU(2)$ gauge theory turns out
to be a double scaling limit of the most simple unitary matrix
model---the so-called one-plaquette or Gross-Witten model \gw.

\subsec{Geometrical engineering of $\cN=2$ theories}

Our strategy will be the following. The $\cN=2$ theory can be
geometrically engineered by taking a suitable limit of type IIB string
theory on a local Calabi-Yau \refs{\klmkv,\klw,\kkv,\kmv}. This local
CY produces directly the Seiberg-Witten curve that encodes the
dynamics of the $\cN=2$ gauge theory.  More precisely the genus zero
topological B-model amplitudes on the local CY capture the
Seiberg-Witten geometry.  The higher genus amplitudes compute the
contributions of the gauge theory to certain gravitational terms of
the form $R^2F^{2g-2}$ \refs{\bcov, \naret}.  We will now engineer a
matrix model that is large $N$ dual to this local CY geometry.  In
particular the planar limit gives the SW geometry and the $1/N$
corrections capture the generation of the corresponding gravitational
terms.

To be specific, let us discuss here the simplest case of pure $SU(2)$
$\cN=2$ Yang-Mills theory. The SW solution is given in terms of the
familiar elliptic curve
\eqn\swc{
w^2=(y^2+u)^2-\L^4,
}
where $u$ is the coordinate on the moduli space, {\it i.e.}\ the vev
of the adjoint $\hf \langle \tr \F ^2\rangle$, and $\L$ the gauge theory
scale, that we will sometimes set conveniently to $\L=1$.

The local CY obtained in the geometric engineering is given by the
algebraic variety \refs{\klw, \kkv , \kmv}
$$
vv'+\L^2\(z+{1\over z}\) + 2(y^2 + u) = 0
$$
with $z\in\C^*$, {\it i.e.}\ an invertible variable. After reducing
over the $(v,v')$-plane the associated Riemann surface is
\eqn\swcurve{
\L^2\(z+{1\over z}\) + 2(y^2 + u) = 0.
}
Since $z\not=0$ we can multiply by $z$ and substituting $w=\L^2 z+
y^2+u$ to bring the curve in the form \swc.

Furthermore, the reduction of the holomorphic three-form on the CY
gives directly the SW differential $ydz/z$.  The prepotential $\cF(u)$
is then obtained by computing the periods of this meromorphic one-form
along the $A$-cycle and $B$-cycle of the elliptic curve \swc. Note
that there are four branch points at $y=\pm\sqrt{-u\pm\L^2}$. In the
classical limit $\L\to 0$ they coalesce pairwise in two double points
$y=\pm\sqrt {-u}$. The moduli space contains two singularities at
$u=\pm 1$ where monopoles respectively dyons become massless.

Since $z$ is a $\C^*$ variable, it makes sense to write $z=e^{ix}$
with the variable $x$ periodic modulo $2\pi$, and reexpress the
original local CY and the resulting curve \swcurve\ as
$$
\L^2 \cos x + y^2+u=0,
$$
with SW differential $ydx$.
This way of writing the equation suggests a relation to a matrix model
with some suitable potential $W(x)$. In fact, since $x$ is now a
periodic variable this suggest a unitary matrix model where $z=e^{ix}$
will get interpreted as the eigenvalue of a unitary matrix $U$.

\subsec{Unitary matrix models}

Unitary matrix models are defined as integrals over the group manifold
$U(N)$ of the form
\eqn\unit{
Z = {1\over {\rm Vol}(U(N))} \int_{U(N)} \!\!\! dU \cdot \exp
(-{1\over g_s}\Tr\,W(U)),
}
where $dU$ is the Haar measure. As in the hermitian matrix models, one
can diagonalize $U$ and express everything in integrals over its
eigenvalues
$$
U \sim {\rm diag}(e^{i\a_1},\ldots,e^{i\a_N}),
$$
where the $\a_i$ are periodic variables, giving
$$
Z = \int \prod_I d\a_I\cdot \prod_{I<J} \sin^2\({\a_I-\a_J \over 2}\)
\exp(-{1\over g_s} \sum_I W(\a_I))
$$

Note that such a unitary matrix model can be viewed as a special case
of a hermitian model by writing $U=e^{i\F}$ with $\F$ a `compactified'
hermitian matrix, {\it i.e.}\ a matrix with a periodic spectrum $\F
\sim \F + 2\pi$. Such a periodicity is achieved by adding multiples of
$2\pi$ to the eigenvalues $\a_I$ of $\F$. This addition of multiple
images is the familiar way to compactify transverse directions for
D-branes or for matrix models in M-theory \taylor. For example, in
this way the Vandermonde determinants in the measure become after
regularization
$$
\prod_{n\in \Z} (\a_I-\a_J+2\pi n) = \sin\({\a_I-\a_J\over 2}\).
$$

The unitary matrix model describes a collection of particles in a
potential $W(\a)$ on the unit circle interacting through a Coulomb
potential. The equation of motion of the unitary model is
$$
W'(\a_I)-2g_s \sum_J \cot\({\a_I-\a_J\over 2}\)=0
$$
The large $N$ solution proceeds exactly as in the uncompactified
case. One introduces again a resolvent, this time defined as
$$
\w(x)=-{1\over N} \sum_I \cot\({x-\a_I\over 2}\),
$$
that satisfies a quadratic loop equation that can be derived in
exactly the same way as in the hermitian case.
In the limit $N\to\infty$ with $g_s N=\mu$ fixed this loop equation
takes the following familiar form, when written in terms of the
variable $y=W'(x)+2\m\,\w(x)$,
\eqn\ucurve{
y^2-W'(x)^2+ 4\m f(x)=0.
}
The quantum correction $f(x)$ in this unitary case is given by the
expression
\eqn\uf{
f(x) = {1\over N} \sum_I \(W'(x)-W'(\a_I)\)
\cot\({x-\a_I\over 2}\)
}

\subsec{The Gross-Witten model}

Our candidate unitary matrix model will be a much studied one, namely
the so-called Gross-Witten model \gw\ with potential $W(\a) \sim
\cos\a$. This model was originally introduced as a lattice discretization
of two-dimensional (non-supersymmetric) Yang-Mills theory. In such a
lattice model to each plaquette with holonomy $U$ around the edge one
associates the Wilson action (in other contexts known as the Toda
potential)
$$
S(U)={\e \over 2g_s}\Tr(U+U^{-1})={\e\over g_s} \Tr\,\cos(\F)
$$
Here $\F$ can be thought of as the lattice approximation to the gauge
field strength $F_{\mu\nu}$, and in the limit $\F\to 0$ this gives the
quadratic Yang-Mills action $\Tr\,\F^2$. (The parameter $\e$ we
introduce for convenience. It can of course be absorbed by rescaling
$g_s=g^2_{YM}$). In a general lattice model one integrates over a
collection of plaquettes, but in two dimensions a single plaquette
suffices to compute for instance a Wilson loop action.

The GW model has two critical points $$ W'(\a)=-\e\sin(\a)=0 $$ at
$a_1=0$ and $a_2=\pi$. Note that for real and positive $\e$ (the case
relevant for Yang-Mills theory) the second point is an unstable
critical point. But that issue is irrelevant for the holomorphic
matrix models that we are considering here. The parameter $\e$ can be
complex, and our eigenvalues $e^{i\a}$ are allowed to move off the
unit circle into the punctured complex plane $\C^*$. In fact,
following our general philosophy, we will consider the perturbative
expansion of the matrix integral \unit\ around the saddle point where
$N_1$ eigenvalues are at the first critical point $a_1$ and $N_2 = N -
N_1$ are at the second point $a_2$. So we are dealing with a two-cut,
meta-stable solution to the matrix integral. These two cuts introduce
a second parameter, besides the overall 't Hooft coupling $\mu=g_sN$,
namely the relative filling fraction
$$
\nu ={(N_1-N_2)/N}.
$$
If we introduce the separate 't Hooft couplings for the two
critical points
$$
\mu_1=g_sN_1,\qquad \mu_2 = g_s N_2,
$$
then the difference of these couplings is related to the filling fraction
$$
\mu'=\mu_1-\mu_2= g_s(N_1-N_2)=\mu\cdot \nu.
$$
We will be interested in computing the planar limit of the free energy
$\cF$ as a function of the two coupling $\mu_1$ and $\mu_2$, or
equivalently as a function of the 't Hooft coupling $\mu$ and the
filling fraction $\nu$. Although $\nu$ takes values in the interval
$[-1,1]$ the final result will turn out to be a holomorphic function
of $\nu$.

The planar limit can be computed solving the loop equation \ucurve. To
this end we have to compute the quantum correction $f(x)$ defined in
\uf. For our choice of potential $W(x)=\e\cos x$ this becomes the following
average over the eigenvalues
\eqn\ff{
\eqalign{
f(x)= &-{1\over N} \sum_I \e\(\sin x-\sin \a_I\)\cot({x-\a_I\over 2}) \cr
= & -{1\over N} \sum_I \e \(\cos x+ \cos \a_I\)\cr
= & -\e (\cos x+u) \cr
}}
Here the constant $u$ is defined as the average
$$
u = {1\over N}\sum_I \cos\a_I.
$$
In the semi-classical approximation $g_s\to 0$ we have
$$
u \approx {N_1-N_2\over N}=\nu.
$$
since there are $N_1$ eigenvalues at the critical point $\a_I=0$ and
$N_2$ eigenvalues at $\a_I=\pi$, that contribute respectively $+1$ and
$-1$ to the average of $\cos\a_I$.

Inserting our expression for $f(x)$ into \ucurve\ gives the spectral
curve
\eqn\bloop{
y^2-\e^2 \sin^2x+ 4\m \e(\cos x+u)=0.
}
We see that the original double points $a_1=0$ and $a_2=\pi$ now split
up in four branch points $a_1^\pm,a_2^\pm$. The two branch cuts
$$
A_1=[a_1^-,a_1^+],\qquad A_2=[a_2^-,a_2^+]
$$
describe the condensation of eigenvalues around the two critical
points.  The discontinuities in $y$ give the eigenvalue density
$\r(\a)$.  Integrating the one-form $ydx$ around the branch cuts $A_i$
gives the filling fractions $N_i/N$. These relations allow one to
express the filling fraction, or more precisely the relative coupling
$\mu'=\mu\,\nu$, in terms of the period
$$
\mu'= {1\over 2\pi i} \oint_{A} y(x) dx
$$
where $A$ is a one-cycle homologous to $A_1-A_2$.  This equation
describes the exact relation between the variables $\mu'$ and $u$.
Integrating the same differential over the conjugated $B$-cycle
encircling the cut $[\a^{(1)}_+,\a^{(2)}_-]$ computes the variation of
the free energy $\cF$ under a change of the relative number of
eigenvalues in the two cuts,
$$
{\d \cF \over \d \mu'} = \int_B y(x) dx.
$$

Note that the GW solution assumes that all eigenvalues center around
the stable vacuum $a_1=0$, in which case $N_2=0$ and $\nu=1$. In fact
we claim that for this vacuum we have exactly $u=1$, so that the
spectral curve is given by
$$
y^2-\e^2 \sin^2x+ 4\m \e(\cos x+1)=0.
$$
Indeed, since one only fills the stable critical point $a_1=0$ with
eigenvalues, only this double point will get resolved into two branch
points. The second critical point $a_2=\pi$ remains unresolved.
Inserting $u=1$ gives the GW solution for the large $N$ eigenvalue
density that can be written as in \gw
$$
\rho(\a)\sim \cos\({\a\over 2}\) \sqrt{{\m\over 2\e}-\sin^2\({\a\over 2}\)}.
$$

\subsec{Double scaling limit}

Now we will take a double scaling limit of the GW model to obtain the
SW solution relevant for $\cN=2$ supersymmetric gauge theory. In this
limit we will send $N\to\infty$ and at the same time $\e\to 0$ and
$\nu\to 0$, keeping $\e N_i$, $\nu/\e$ and $g_s$ fixed---or,
equivalently, we will send the 't Hooft coupling $\mu\to\infty$
keeping the difference of the two couplings $\mu_1$ and $\mu_2$
$$
\mu'=g_s(N_1-N_2)=\mu\,\nu
$$
fixed. In this limit the absolute difference in eigenvalues $N_1-N_2$
remains finite, but $N_1$ and $N_2$ become both infinite, and
therefore the relative filling fraction $\nu=(N_1-N_2)/N$ goes to
zero.

After rescaling $y$ appropriately, the spectral curve reduces in this
limit exactly to the SW curve (with $\L=1$)
$$
y^2+\cos x + u = 0.
$$
Note that the double scaled curve depends only on a single parameter
$u$ that at weak coupling could be identified with the filling
fraction $\nu=(N_1-N_2)/N$ of the matrix model.  Of course the limit
we are taking is at a strong coupling point and the relation between
$u$ and matrix model modulus $\mu'$ is more complicated, as discussed
above.

As we have already mentioned, the prepotential is now computed by the
periods of the differential $ydx=ydz/z$ along the $A$ and
$B$ cycles. This allows us to identify the SW periods as
\eqn\aad{
\eqalign{
a &= \oint_A y(x)dx = \mu'\cr
a_D &= \oint_B y(x)dx = {\d \cF \over \d\mu'}\cr
}}

{}From the original four branch points $a_1^\pm,a_2^\pm$ obtained in
resolving the two double points $a_1,a_2$, our double scaling limit
takes the branch points $a_1^-,a_2^+$ to infinity while keeping
$a_1^+,a_2^-$ at finite distance. This leaves two homology one-cycles:
the $B$-cycle that runs around the cut $[a_1^+,a_2^-]$ and the dual
$A$-cycle that is homologous to $A_1-A_2$. This behaviour of the
branch points is exactly the behaviour in the double scaling limit one
takes in the old matrix models \refs{\mm,\mmm}, as we also noted in
\dv. For example, the $(2,3)$ critical point of the one-matrix model
was obtained starting from a curve of the form $y^2=x^6+\ldots$ and
the double scaling limit got rid of the all monomials with power more
than $x^3$ giving an equation of the form $y^2=x^3+\ldots$ (Note that
these branch points in the $x$-plane should not be confused with the
branch points in the $y$-plane that are relevant for the SW solution.)

The double scaling is very analogous to the limit that was used in
the A-model topological string in \amv. In fact, when the scaling
limit is embedded in type II string theory, the resulting CY geometry
based on \bloop\ will have RR flux through the compact cycles $A_1$
and $A_2$. It is crucial that the remaining compact cycle $B$ does not
carry any Ramond flux. We are thus engineering a large $N$ dual of a
geometry without fluxes.

The GW model has a famous third order phase transition at (in our
convention) $\mu/\e=2$. This signals a transition of the eigenvalue
distribution in which the single cut changes topology and starts to
cover the whole unit circle. After the GW phase transition the
eigenvalue distribution is given by $\rho(\a) \sim \cos
\a+\m/2$. Geometrically speaking, in that phase all
four branch points are on top of each other.

This phase transition is however not relevant in our model. First of
all, we are studying a more general question by considering a
stationary phase approximation around a meta-stable state with two
clusters of eigenvalues and consequently have to work with a
two-dimensional phase diagram $(\m,\n)$. As we argued the GW solution
puts $\n=1$ and that is very far away from our double scaling limit in
which $\nu$ tends to zero. Indeed in our limit the number of
eigenvalues in the two cuts is roughly equal.  Secondly, we are
dealing with an holomorphic object, and holomorphy excludes any phase
transitions, one can just go around the singularity. The GW phase
transition is just a (very special) real slice of our complex phase
diagram.

Finally it would be interesting to connect this approach to the
beautiful semi-classical computation of the SW solution and its
gravitational counterparts in \nekrasov. That computation was inspired
by matrix integrals appearing in D-branes formulas.

\subsec{Generalizations for general groups}

It is not difficult to guess how the SW solution for gauge group
$SU(n)$ can be engineered. In this case the curve associated to the
local CY is of the form
$$
\cos x + P_n(y)=0,
$$
with $P_n(y)$ a polynomial of degree $n$.
More generally we can consider a chain of $U(n_i)$ gauge theories
with bifundamental matter, for which the corresponding curve
has been obtained from the M5 brane viewpoint in \witf\ and from
the viewpoint of geometric engineering in \kmv .  This will
give rise to a curve of the form
$$
F(e^{ix}, y)=0
$$
where $F$ is a polynomial in $e^{\pm ix}$ and $y$.  In particular
if we consider the rank of all the gauge groups to be equal to $n$,
then $F$ is a polynomial in $y$ of degree $n$.  Moreover the difference
in power of $e^{ix}$ between the highest and lowest powers is the
number of $U(n)$ gauge groups plus 1.
 As we will discuss in greater
detail in the next section in the context
of hermitian multi-matrix models such curves are typically produced by a
multi-matrix model consisting of $n-1$ matrices.
The choice of the coefficients in $F$ will be related
to the choice of the action and some suitable double scaling limit,
as we studied in the context of $SU(2)$ gauge theory here.

\subsec{Connections with A-model topological strings}

It is natural to ask if there is a connection with A-model
topological strings, and in particular for A-models
on local toric CY.  As was demonstrated in \refs{\hv,\hiv}\
the B-model mirrors are given by
$$vv'+F(e^{ix},e^{iy})=0$$
for some $F$.  This is analogous to an infinite matrix model version
of the unitary matrix models, as follows from our discussion above.
In this case the A-model has a gauge theory dual involving certain
correlations functions of Chern-Simons theory \refs{\amv, \dfg}. It
would be interesting to connect these matrix models directly with the
Chern-Simons gauge theory computation, thus completing the circle of
ideas.  There are some hints that this idea indeed works
\marinor .

\newsec{Quiver matrix models}

We will now turn to a related generalization of \dv\ where we will
connect superpotential computations of quiver gauge theories to
multi-matrix models.

\subsec{$\cN=1$ quiver gauge theories and topological strings}

We will restrict our discussion here to the $ADE$ quivers, in
particular the $A_r$ case, although one can also include the affine
quivers based on the extended Dynkin diagrams $\widehat A\widehat
D\widehat E$. Let $r$ denote the rank of the quiver $G$ and consider a
partition
$$
N=N_1+\ldots+ N_r.
$$
In the associated $\cN=2$ quiver gauge theory we assign to each of the
$r$ vertices $v_i$ of the Dynkin diagram of $G$ a $U(N_i)$ gauge field
and to links connecting vertices $v_i$ and $v_j$ we associate
bifundamentals $Q_{ij}$ transforming in the representation
$({N}_i,\overline{N}_j)$ with a hermiticity condition $Q^\dagger_{ij}=
Q_{ji}$.

For such a gauge theory we can write a general tree-level
superpotential
\eqn\superpot{
W(\F,Q)=\sum_{i,j} s_{ij}\Tr\,Q_{ij}\F_j
Q_{ji}+\sum_i \Tr\,W_i(\F_i), }
with $s_{ij}=-s_{ji}=1$ (for some ordering $i<j$), if the vertices
$v_i$ and $v_j$ are linked in the Dynkin diagram (we will write this
relation also as $\langle i,j\rangle$), and $s_{ij}=0$ otherwise.  Here
the first term is the standard superpotential of the $\cN=2$ theory
with bifundamental matter. The additional potentials $W_i(\F_i)$ are
introduced to break the supersymmetry down to $\cN=1$.

Within type II string theory these quiver gauge theories are obtained
by wrapping D5-branes over a particular CY geometry that is a
fibration of the corresponding $ADE$ singularity over the complex
plane \ckv. This geometry contains $r$ intersecting
$\P^1$'s. According to \refs{\ckv,\cfikv} in the large $N$ limit the
geometry undergoes a transition to a deformed geometry where these
$\P^1$'s are blown down and a number of $S^3$'s with RR flux are
``blown up.''  The corresponding smooth CY geometry gives a dual
description of the gauge theory system.

In the context of B-model topological strings, the deformed CY
geometry is dual to a two-dimensional large $N$ gauge system, obtained
from a collection of B-branes wrapped on the intersecting
$\P^1$'s. The world-volume theory of these branes consists of open
topological strings. So each $\P^1$ gives rise to a two-dimensional
field theory with Lagrangian \kkl
\eqn\act{S(\F)= {1\over g_s}\int_{\P^1} \Tr\Bigl(
\F^1_i{\overline D_A}\F^0_i +W_i(\F^0_i) \omega \Bigr).}
where $\omega$
is some volume form on $\P^1$, and $\F^0_i$ and $\F^1_i$
are adjoint
fields of respectively spin 0 and spin 1 coupled to an $U(N_i)$
holomorphic gauge field. Here we included the effect of the
superpotential $W_i(\F_0)$. The open topological strings connecting
different $\P^1$'s give as physical fields the bifundamentals
$Q_{ij}$. Since the different $\P^1$'s intersect in points, the action
of these bifundamental scalar fields localizes to the intersection
point $x$ and is given by
$$
S(Q) = \sum_{\langle i,j \rangle} \sum_{x\in \P^1_i \cap \P^1_j}
\Tr \(Q_{ij}(x)\F_j^0(x) Q_{ji}(x) - Q_{ji}(x)\F_i^0(x) Q_{ij}(x)\).
$$
(Compare the similar computation for the coupling of open topological
strings connecting Lagrangians A-branes intersection along one-dimensional
curves in \ovknot.)

As in \dv\ one can see that in the end this two-dimensional
topological field theory can be completely reduced to the zero modes
of the fields $\F^0_i(x)=\F_i$ and $Q_{ij}$. Thereby the path-integral
reduces to the ``quiver matrix integral''
\eqn\qmm{
Z= \int \prod_i d\F_i \prod_{\langle i,j\rangle}
dQ_{ij}
\exp\Bigl(-{1\over g_s}{\rm Tr}\,W(\F,Q) \Bigr).
}
where $\F_i$ (with $i=1,\ldots,r$) is an $N_i\times N_i$ hermitian
matrix and for every linked indices $\langle i,j\rangle$ the variable
$Q_{ij}$ is a $N_i\times N_j$ rectangular matrix, satisfying
$Q^\dagger_{ij}=Q_{ji}$.

The generalization of the conjecture in \dv\ will now identify the
free energy of the large $N$ quiver matrix model, for given filling
fractions of the saddle points, with the closed topological string
partition function in the corresponding deformed CY geometry, and this
in turn with the effective superpotential of the quiver gauge theory.

\subsec{Saddle points and dual CY geometry}

The saddle points of the quiver superpotential have been discussed
extensively in \refs{\ckv,\cfikv,\ot} following the mathematical
literature. The eigenvalues $x$ of the adjoints $\F_i$ have to satisfy
a series of equations: one for every positive roots $\a_k$ of $G$. If
that root is expressed in the simple roots $e_i$ as
$$
\a_k = \sum_i n^i_k e_i,
$$
hen the associated condition reads
\eqn\root{
\sum_i n^i_k W'_i(x)=0.
}
The saddle points can be labeled as $x_{a,k}$ with $a=1,\ldots,d$ and
$d$ the maximal degree occurring in \root. If such a critical point
appears with multiplicity $N_{a,k}$ then the total number of
eigenvalues of the matrix $\F_i$ in this saddle point is $N_{a,k}
\cdot n^i_k$. A general saddle point is therefore parametrized by the
filling fractions $\nu_{a,k} = {N_{a,k}/N}.$ We will consider the
matrix integral in the limit where both the $N_{a,k}$ and $N$ tend to
infinity keeping the filling fractions and the 't Hooft coupling
finite.

In the case of an $A_r$ quiver there is a more straightforward
description of the saddle points \refs{\ckv,\cfikv,\ot}. Introduce
the $r+1$ potentials
$$
t_0(x)=0,\qquad t_i(x)=\sum_{j=1}^i W_j'(x),\ \ \ i=1,\ldots,r.
$$
The the associated rational planar curves
$$
y-t_i(x)=0
$$
intersect in various double points given by
$$
t_j(x)-t_i(x)=\sum_{k=i+1}^j W_k'(x)=0.
$$
The saddle points of the $A_r$ quiver matrix potential correspond
exactly to these double points.

In this $A_r$ case the original singular CY geometry is given by
$$
uv+\prod_{i=0}^{r}\(y-t_i(x)\)=0
$$
which after reduction over $u,v$ gives precisely this collection of
nodal curves. After the deformation the corresponding smooth Riemann
surface is given by
\eqn\ar{
\prod_{i=0}^{r}\(y-t_i(x)\)+ f(x,y)=0.
}
for a suitable normalizable quantum deformation $f(x,y)$. Again every
double point gets resolved into two branch points. The resulting
quantum curve is now an $r+1$ fold cover of the $x$-plane. By moving
around in the $x$-plane these sheets will be exchanged through Weyl
reflections acting on the parameters $t_i$.

The analogues of the meromorphic one-form are constructed by reducing
the holomorphic three-form of the local CY over various cycles and for
$A_r$ have the following description \refs{\ckv,\cfikv}.
Write the curve \ar\ in the
factorized form
$$
\prod_{i=0}^r \(y-a_i(x)\)=0.
$$
Then we have a basis of $r-1$ meromorphic one-forms $\eta_1,\ldots,
\eta_{r-1}$ that is
in one-to-one correspondence with a basis of positive roots for
$A_{r-1}$ given by
\eqn\onef{
\eta_i = \(a_{i+1}(x)-a_i(x)\)dx.
}
In the undeformed case, with $a_i(x)=t_i(x)$, this gives $\eta_i=
W_i'(x)dx=dW_i$. Different choices of positive roots correspond to Weyl
reflections, which are in fact the generalized Seiberg dualities of
gauge theories (which should also have some direct interpretation as
dualities of the matrix models).  Note that in the case of $A_1$ we
have
$$
(y-a)(y+a)=0,\qquad a^2=W'(x)^2+f(x)
$$
and this gives the usual definition $\eta=\pm y dx.$

We will argue that this process of smoothing out the singular curve
including the meromorphic differentials is exactly described by the
large $N$ dynamics of the quiver matrix integral.

\subsec{$ADE$ matrix models}

Quite remarkably it turns out that the quiver matrix integrals \qmm\
(up to some minor details) have already been studied in the context of
the ``old matrix models.'' They have been used to describe the
coupling of $ADE$ conformal minimal models to two-dimensional gravity
by Kostov
\kostov, see also the reviews \refs{\ki,\kii}, and they have naturally
emerged in the study of matrix models and integrable systems in the
work of the ITEP group \kmmms.  We will follow closely these works in
presenting the main results, leaving the details to the literature.

First of all, one can immediately integrate out the bifundamental
fields $Q_{ij}$ in the quiver matrix integral to give an effective
interaction between the adjoint fields $\F_i$ and $\F_j$
$$
\det\(\F_i\otimes {\bf 1}-{\bf 1}\otimes \F_j\)^{-1}
$$
After expressing everything in terms of the eigenvalues of the
remaining hermitian matrices $\F_i$
$$
\F_i \sim {\rm diag}\(\l_{i,1},\ldots,\l_{i,N_i}\)
$$
the quiver matrix integral reduces to
$$
Z = \int \prod_{i,I} d\l_{i,I} \cdot {\displaystyle
\prod_{i,I<J}\(\l_{i,I}-\l_{i,J}\)^2 \over \displaystyle
\prod_{i<j,I,J}\(\l_{i,I}-\l_{j,J}\)^{|s_{ij}|}} \exp
\Bigl({-{1\over g_s} \sum_{i,I} W_i(\l_{i,I})}\Bigr).
$$
Note that if we introduce the Cartan matrix of $G$
$$
C_{ij}=2\delta_{ij} -|s_{ij}|=e_i\cdot e_j,
$$
then the quiver eigenvalue measure, that generalizes the usual
``fermionic'' Vandermonde determinants of the one matrix model
$$
\Delta(\l)^2= \prod_{I<J} \(\l_{I}-\l_{J}\)^2,
$$
can now be written as
$$
\prod_{(j,J)\not=(i,I)} \!\! \(\l_{i,I}-\l_{j,J}\)^{e_i\cdot e_j/2}.
$$
The resemblance to a correlation function of vertex operators is not
accidental---it was in fact the main motivation to study these kind of
matrix models in \kmmms since it allows one to express the partition
function as a particular state in a two-dimensional chiral CFT, to be
more precise the level one realization of the corresponding $ADE$
current algebra.

In the large $N$ limit this many-flavor Dyson gas of eigenvalues will
spread out in cuts around the saddle points and will form a continuum
of eigenvalues described by a series of densities functions
$$
\rho_i(x)={1\over N} \sum_{I} \delta(x-\l_{i,I}).
$$
The solution of the model proceeds again through the resolvents or
loop operators of the matrices $\F_i$
$$
\w_i(x)= {1\over N}\Tr\({1\over x-\F_i}\) =
{1\over N} \sum_{I} {1\over x-\l_{i,I}}.
$$
The jump of $\w_i(x)$ across a branch cut measure the eigenvalue
density $\r_i(x)$.

In fact, it is natural to work with a closely related object---the
derivative of the matrix model action $S$ with respect to an
eigenvalue of type $i$ evaluated at a general position $x$ in the
complex plane (away from the cuts) as it interacts with all the other
eigenvalues,
\eqn\yi{
y_i(x)=g_s \,\d_i S =  W_i'(x) -2 \mu \,C^{ij} \w_j(x).
}
As we already mentioned in \dv\ for a multi-matrix model we want to
identify the one-forms $\eta_i$ \onef\ coming from the local CY
geometry with the expressions $y_i(x)dx$, up to a possible change of
basis.  The most powerful and general technique that can be used to
relate the CY geometry to the matrix model are the loop equations.

\subsec{Loop equations and collective fields}

Before we discuss the loop equations of the multi-matrix models, let
us first rewrite the solution of the one-matrix model as used in \dv\
in a more suggestive form, that is actually a standard technique in
matrix model technology. Here we found among others the reviews
\refs{\ki,\kii,\morozov} very helpful.

The resolvent $\w(x)$ has a natural interpretation as a loop operator.
More precisely, the inverse Laplace transform
$$
\int {dx \over 2\pi} \, e^{ix\ell}\, \w(x) =
\Tr\(e^{\ell \F}\)
$$
is the zero-dimensional analogue of the Wilson loop. The non-linear
all-genus loop equation is usually written in terms of $\w(x)$ as
\eqn\loopeqn{
\oint_\cC {dz \over 2\pi i} {W'(z) \over x-z}\< \w(z) \> =
\mu \<\w(x)^2\>
}
where $\<\cdots\>$ indicates an expectation value within the matrix
integral. The contour $\cC$ encircles all the cuts but {\it not} the
point $x$. This equation is supplemented with the boundary condition
$\<\w(x)\> \sim 1/x$ at infinity. The loop equation acts as a
Schwinger-Dyson equation of the matrix model. It gives a recursive
relation to solve for the loop operator and the free energy. In the
planar limit we have large $N$ factorization $\<\w(x)^2\>=\<\w(x)\>^2$
and the loop equation becomes algebraic.

Loop operators are closely connected to collective fields.  By
integrating out the angular variables the individual eigenvalues start
to behave as fermions, and the collective field is essentially
constructed by bosonization of these fermion fields. In
\dv\ we have already speculated that this collective field should be
identified with the Kodaira-Spencer field \bcov\ describing the closed
strings moving on the local CY geometry.

For a single matrix model the collective field is defined as the
chiral two-dimensional scalar field
$$
\v(x)=W(x) -2g_s \sum_I \log(x-\l_I)
$$
It clearly satisfies $\d\v(x)=y(x)$ with
$$
y = W'(x) - 2g_s \sum_I {1\over x-\l_I}=W'(x) -2\mu\w(x)
$$
So in view of \yi\ we can identify the function $\v(x)$ with the
action $S(x)$ of a single eigenvalue as a function of its position $x$
in the complex plane in the presence of the gas of other eigenvalues
$\l_1,\ldots,\l_N$. The function $\v(x)$ is multi-valued in the
$x$-plane. It has branch cuts around which it changes sign. It is
therefore only properly defined on the double cover
\eqn\cover{
y^2-W'(x)^2+f(x)=0.
}
On this Riemann surface $\v(x)$ has quantized periods around the
$A$-cycles, given by the filling numbers $\mu_i=g_s N_i$.  Since it is
a chiral field the periods around the dual $B$-cycles are not
independent and expressed by the special geometry relations
as $\d\cF/\d\mu_i$.

Note that if we work with a general, not necessarily polynomial,
superpotential
$$
W(x) = \sum_{n\geq 0} t_{n} x^n,
$$
then the expectation value of the field $\d\v(x)$ inserted in the
matrix integral can be represented by a linear differential operator
in the couplings $t_n$ acting on the partition function. For example,
$$
\langle \d\v(x)\rangle =
\Bigl(\sum_{n>0} nt_{n} x^{n-1} -2 g_s^2 \sum_{n\geq 0}
x^{-n-1} {\d\over \d t_{n}}\Bigr) Z,
$$
and similarly for multi-point functions. (Here we used that the
derivative $\d/\d t_n$ brings down a factor ${1\over g_s}\Tr\,\F^n$.)

With this notation there is an elegant way to write the loop
equations. Introduce the holomorphic stress-tensor
$$
T(x) = (\d\v)^2=\sum_n L_{n} x^{-n-2}
$$
Then the all-genus loop equation \loopeqn\ of the one-matrix model can
be rewritten in the suggestive form
\eqn\stress{
\oint_{\cC} {dz \over 2\pi i} {1\over x-z} \<T(z)\>  = 0,
}
That is, the expectation value $\langle T(x) \rangle$ has no
singular terms if $x\to 0$. Therefore we can also express
\stress\ equivalently as the Virasoro constraints \refs{\dvv,\kawai}
$$
L_nZ=0,\qquad n\geq -1.
$$
The derivation of the constraints in the matrix model is completely
standard---it simply expresses the Ward identities following from the
invariance under infinitesimal reparametrization of $\F \to \F +
\epsilon \F^{n+1}$ of the matrix variable $\F$.

In the planar limit we can substitute the classical values for
$\d\v(x)=y$ in $T(x)$ and then equation \stress\ is a consequence of
\cover\ that can now be written as
$$
T(x)=W'(x)^2-f(x),
$$
which shows that $T(x)$ is indeed regular (even polynomial) at $x=0$.

\subsec{Quiver theories and $W$-constraints}

The large $N$ solution of the quiver matrix integral \qmm\ now
proceeds along similar lines \refs{\kostov,\kmmms}.  One introduces
$r$ scalar fields $\v_i(x)$ through the one-forms \yi\ as
$$
y_i(x)dx=\d\v_i(x)
$$
One can then show that the multi-valued fields $\v_i(x)$ are actually
the values of one single-valued field $\v(x)$ (essentially the full
matrix model action) on a $r+1$ branched cover of the complex
$x$-plane. This branched cover is the spectral curve associated to the
quiver matrix integral, and turns out to be given by \ar\ in the $A_r$
case.

The general derivation of the curve proceeds through generalized loop
equations. For these multi-matrix model we do not only have the
Virasoro constraints, expressing reparametrization invariance in the
matrix variables $\F_i$. There are also higher order relations
\refs{\dvv,\kawai}.  The full set of loop equations are
obtained by showing that the partition function $Z$ satisfies a set of
$W$-constraints, labeled by the Casimirs of the corresponding $ADE$
Lie algebra, which contains the Virasoro constraints. These constraints
take the form
$$
\oint_\cC {dz \over 2\pi i} {1 \over x-z}\< \cW^{(s)}(z) \> = 0,
$$
where $\cW^{(s)}(x)$ is a spin $s$ current in the $W$-algebra. When
expressed in modes these equations take the form
$$
\cW^{(s)}_n\cdot Z=0,\qquad n\geq 1-s.
$$
In the case of $A_r$ there is leading spin $r+1$ current that with a
suitable basis of vectors $\v_0,\ldots,v_r$ can be written as
$$
\cW^{(r+1)}(x) \sim \prod_{i=0}^r (v_i\cdot \d\v)+\ldots
$$
We claim that in the planar limit this loop equation translates
directly into the curve \ar.

To be completely explicit let us give some more detail for the
simplest case of $A_2$.  Here we have two matrices $\F_1,\F_2$
with potentials $W_i(\F_1)$ and $W_2(\F_2)$. The classical singular
curve is after a shift in $y$ given by
$$
\(y-t_1(x)\)\(y-t_2(x)\)\(y-t_3(x)\)=0
$$
with
$$
t_1=-(2W_1'+W_2')/3,\quad
t_2=(W_1'-W_2')/3,\quad
t_3=(W_1'+2W_2')/3,
$$
all polynomials in $x$. To find the quantum curve we introduce the
resolvents
$$
w_1(x)=\sum_I {1\over x - \l_{1,I}},\qquad
w_2(x)=\sum_I {1\over x - \l_{2,I}},
$$
and the one-forms $y_i(x)dx$
$$
y_1 = W_1' - \mu(2\w_1-\w_2), \qquad
y_2 = W_2' - \mu(2\w_2-\w_1).
$$
We now claim that the quantum curve is given by
$$
\(y-a_1(x)\)\(y-a_2(x)\)\(y-a_3(x)\)=0,
$$
where the functions $a_i(x)$ are no longer polynomials, but instead
are defined as
$$
a_1=t_1+\m\,\w_1,\quad
a_2=t_2-\m(\w_1-\w_2),\quad
a_3=t_3-\m\,\w_2.
$$
With this choice we have, as claimed before,
$$
a_2-a_1=y_1,\qquad a_3-a_2=y_2.
$$
Now after some algebra, expanding out terms like
$\w_i(x)^3$, one verifies that indeed
$$
\eqalign{
& \(y-a_1(x)\)\(y-a_2(x)\)\(y-a_3(x)\)= \cr &
\qquad\qquad \(y-t_1(x)\)\(y-t_2(x)\)\(y-t_3(x)\)
 +f(x)\, y+g(x)=0 \cr
}
$$
with $f(x)$ and $g(x)$ polynomials.

\bigskip
\centerline{\bf Acknowledgements}

We would like to thank J.~de Boer, M.~Mari\~no, and E.~Verlinde for
discussions.  The research of R.D. is partly supported by FOM and the
CMPA grant of the University of Amsterdam, C.V. is partly supported by
NSF grants PHY-9802709 and DMS-0074329.

\listrefs

\bye